\documentclass[11pt]{article}

\title{\texttt{langcc}: A Next-Generation Compiler Compiler}
\author{Joe Zimmerman \vspace*{-0.1em} \\ \small{\texttt{jzim@cs.stanford.edu}}}
\date{}

\usepackage{amsfonts}
\usepackage{amsmath}
\usepackage{amssymb}
\usepackage{amsthm}
\usepackage{array}
\usepackage{enumerate}
\usepackage{enumitem}
\usepackage[left=1in,right=1in,top=1in,bottom=1in,columnsep=0.25in]{geometry}
\usepackage[hidelinks]{hyperref}

\setlist{listparindent=\parindent,parsep=0pt}

\begin{document} 

\maketitle

\begin{abstract}
    Traditionally, parsing has been a laborious and error-prone component of compiler development,
    and most parsers for full industrial programming languages are still written by hand.
    The author \cite{Zim22} shows that automatic parser generation can be practical,
    via a number of new innovations upon the standard~LR~paradigm of Knuth~et~al.
    With this methodology,
    we can automatically generate efficient parsers for virtually all languages
    that are intuitively ``easy to parse''.
    This includes Golang 1.17.8 and Python 3.9.12,
    for which our generated parsers are, respectively, 1.2x and 4.3x faster
    than the standard parsers.
    This document is a companion technical report which describes
    the software implementation of that work,
    which is available open-source at~\url{https://github.com/jzimmerman/langcc}.
\end{abstract}

\section{Introduction}

\texttt{langcc} can be used as a replacement for the combination of \texttt{lex}
and \texttt{yacc} (or \texttt{flex} and \texttt{bison}). However, \texttt{langcc} provides many
additional features, including:
\begin{itemize}
    \item Automatic generation of AST data structures, via a standalone datatype
        compiler (\texttt{datacc}).
    \item Full LR parser generation as the default, rather than the more restrictive
        LALR.
    \item Clear presentation of LR conflicts via explicit ``confusing input pairs'',
        rather than opaque shift/reduce errors.
    \item Novel efficiency optimizations for LR automata.
    \item An extension of the LR paradigm to include recursive-descent (RD)
        parsing actions, resulting in significantly smaller and more intuitive
        automata.
    \item An extension of the LR paradigm to include per-symbol attributes,
        which are vital for the efficient implementation of many industrial language
        constructs.
    \item A general transformation for LR grammars (CPS), which significantly
        expands the class of grammars the tool can support.
\end{itemize}
Unlike previous compiler compilers, \texttt{langcc} is efficient and general enough to capture
full industrial programming languages, including
Python 3.9.12 (\texttt{grammars/py.lang}) and Golang 1.17.8
(\texttt{grammars/go.lang}). In both cases, \texttt{langcc} automatically generates
a parser that is faster than the standard library parser for each language
(resp., 1.2x and 4.3x faster).
In fact, the class of grammars supported by \texttt{langcc} is general enough
that the tool is {\em self-hosting}: that is, one can express the ``language
of languages'' in the ``language of languages'' itself, and use \texttt{langcc}
to generate its own compiler front-end. We do this in the canonical
implementation; see the files \texttt{bootstrap.sh} and \texttt{grammars/meta.lang}
in the source repository for more details.

\section{Usage}

\subsection{\texttt{langcc}}

\texttt{langcc} is a standalone command-line tool,
which can be invoked as follows:
\[ \texttt{langcc X.lang gen\_path} \]
to compile a BNF-style language specification (\texttt{X.lang})
to a generated compiler front-end for that language (\texttt{gen\_path/X\_\_gen.hpp}, \texttt{gen\_path/X\_\_gen.cpp}).
The tool automatically generates data structures for the abstract syntax (AST),
a lexer, a parser, and a pretty-printer for terms in the language.
In order to use the generated code, simply include the file \texttt{gen\_path/X\_\_gen.hpp}
and compile and link against the file \texttt{gen\_path/X\_\_gen.cpp}.

\subsection{\texttt{datacc}}

\texttt{datacc} is an internal component which \texttt{langcc} uses to generate many of its
data structures, but which can also be used as a standalone command-line tool.
It can be invoked as follows:
\[ \texttt{datacc X.data gen\_path} \]
to compile a declarative specification of some algebraic datatypes (\texttt{X.data})
to generated C++ code that implements those datatypes
(\texttt{gen\_path/X\_\_data\_gen.hpp}, \texttt{gen\_path/X\_\_data\_gen.cpp}).
\texttt{datacc} supports named product and sum types (with enums as a special case),
as well as datatypes with type parameters.
The generated code includes the (reference-counted) C++ implementations
of the corresponding algebraic datatypes as structs with inheritance,
as well as a number of other features:~e.g.,
functions for testing and downcasting sum types,
functions for substituting fields of product types,
functions for debug-printing elements of each datatype,
and functions for cached value-based SHA-256 hashing of elements of each datatype.

\section{Input language}

The input to \texttt{langcc} is a file \texttt{X.lang},
written in the ``language of languages'' (\texttt{grammars/meta.lang}).
Such an input consists of the following stanzas:
\begin{itemize}
\item \texttt{tokens}: Describes the basic tokens that are to be
    emitted by the lexer.
    A token description is either an {\em opaque} declaration \texttt{X <- e}
    or an {\em alias} declaration \texttt{X <= e},
    where \texttt{e} in either case is (roughly) a regular expression.
    For instance, one can write:
    \vspace*{-0.3em}
    \begin{verbatim}
    int_lit <- `0` | (`1`..`9`) digit*;
    digit <= `0`..`9`;
    op <= `+` | `-` | `*` | `/`;
    \end{verbatim}
    \vspace*{-1.5em}
    Backtick-quoted strings are used to represent
    literal sequences of characters that appear in the tokens.
    Expressions can be parenthesized,
    and many standard regex operators
    (e.g., concatenation, alternation, repetition,
    parenthesization, character ranges)
    are permitted.
    Standard escapes such as \texttt{\textbackslash n} and \texttt{\textbackslash \textbackslash}
    are valid within backtick-quoted strings,
    and the backtick itself can be escaped via \texttt{\textbackslash `};
    however, single and double quotes do not need to be escaped.

    Note that in the example above, \texttt{digit} is an alias,
    and thus can be used in the definition of \texttt{int\_lit}.
    However, \texttt{int\_lit}, as an opaque declaration,
    cannot be used in the definition of other items in the lexer---as opposed to the parser,
    which does permit recursive structure.
    The other key difference between alias and opaque token declarations
    is that opaque tokens will be emitted as such by the lexer,
    and can appear directly in parser expressions
    (e.g., \texttt{Expr.Int <- n:int\_lit});
    while if aliased tokens are emitted by the lexer,
    then the item that is actually emitted is the underlying opaque token expression
    (e.g., in the parser, below, we could write
    \texttt{Expr.Add <- x:Expr `+` y:Expr}, but we could not write
    \texttt{Expr.Binop <- x:Expr op y:Expr}, as \texttt{op} is not an
    opaque token emitted by the lexer).

\item \texttt{lexer}: Describes the procedural operation of the lexer
    as it processes an input string to emit tokens.
    The lexer stanza consists of one or more {\em lexer modes},
    as well as a {\em main mode} declaration.
    For instance, the following is a lexer stanza:
    \begin{verbatim}
    lexer {
        main { body }

        mode body {
            top => { emit; }
            ws_inline => { pass; }
            `\n` => { pass; }
            `//` => { push comment_single; pass; }
            eof => { pop; }
        }

        mode comment_single {
            `\n` => { pop_extract; }
            eof => { pop_extract; }
            _ => { pass; }
        }
    }
    \end{verbatim}
    \vspace*{-1.5em}
    Here, the tokens `\texttt{eof}' and `\texttt{\textbackslash n}' are built-in;
    the underscore `\texttt{\_}' is a wildcard matching an arbitrary unicode
    code point;
    and we assume that the aliases `\texttt{top}', `\texttt{ws\_inline}'
    have been defined in the \texttt{tokens} section.
    The main mode is the one called `\texttt{body}',
    and in this mode:
    \begin{itemize}
        \item If the lexer encounters a toplevel token
            (i.e., a character sequence matching one of
            the opaque constituents of the alias \texttt{top}),
            then it will emit that constituent, and proceed.
        \item If the lexer encounters inline whitespace,
            then it will pass over it and proceed.
        \item If the lexer encounters the newline character,
            then it will pass over it and proceed.
        \item If the lexer encounters the comment start sequence `\texttt{//}',
            then it will push the mode \texttt{comment\_single}
            onto its mode stack,
            pass over the `\texttt{//}',
            and continue processing in the new mode.
            Note that the ordering of the two commands is significant.
            The second command, `\texttt{pass}',
            means that the matched string `\texttt{//}'
            will be processed {\em after} the lexer is already in the new mode
            \texttt{comment\_single}.
        \item If the lexer encounters the end-of-file marker (`\texttt{eof}'),
            then it will pop the main mode off the stack.
            Note that if at any point, the mode stack is empty,
            then the lexer halts;
            and it declares success if and only if this happens at the end of the input---if
            the mode stack is empty prematurely, this is a failure.
            Conversely, if the mode stack is nonempty at end-of-file,
            this is also a lexing failure.
    \end{itemize}
    while, in the mode \texttt{comment\_single}:
    \begin{itemize}
        \item If the lexer encounters either a newline or the end-of-file marker,
            it will extract all of the characters that have been processed
            in the current mode (including those that have been passed over),
            attach the result as extra data (not part of the AST),
            and pop the current mode off of the mode stack.
        \item If the lexer encounters any other character,
            it will pass over it, and proceed.
    \end{itemize}
    Note that there are two possible types of pop commands:
    \begin{itemize}
        \item \texttt{pop\_extract}, which is used above.
            This will extract the characters processed in the current mode,
            and attach the result as extra data (not part of the AST).
        \item \texttt{pop\_emit tok}, where \texttt{tok} names an opaque token.
            This will extract the characters processed in the current mode,
            and emit an instance of the token \texttt{tok} (to be consumed by the parser),
            where the included string contents of \texttt{tok} are the extracted characters.
    \end{itemize}
    The generated lexer is implemented via the standard NFA/DFA subset construction,
    and it will accept the {\em longest} matching substring
    starting at the current point in the input
    (in other words, it will only consider performing an action if the subsequent character
    would lead to a DFA state with no action).
    By construction of the DFA, the action of a compiled lexer is always uniquely determined.
    While some lexer/token definitions may lead to ambiguity,
    this can be detected statically in the subset construction,
    and this will generate an error during \texttt{langcc}'s compilation of \texttt{X.lang}
    rather than producing an ambiguity at lexing time.

\item \texttt{parser}: Describes the context-free grammar which should be used
    by the generated parser,
    defined in terms of the basic lexer tokens.
    For instance, the following is a parser stanza:
    \begin{verbatim}
    parser {
        main { Stmt, Expr }

        prec {
            Expr.BinOp1 assoc_left;
            Expr.BinOp2 assoc_left;
            Expr.UnaryPre prefix;
            Expr.BinOp3 assoc_left;
            Expr.Id Expr.Lit.Int_ Expr.Paren;
        }

        prop { name_strict; }

        Stmt.Assign <- x:Expr[I] _ `=` _ y:Expr;
        Stmt.Expr <- x:Expr;

        Expr.Id[I] <- name:id;
        Expr.Lit.Int_ <- val:int_lit;
        Expr.UnaryPre <- op:#Alt[Neg:`-`] x:Expr;
        Expr.BinOp1 <- x:Expr _ op:(Add:`+` | Sub:`-`) _ y:Expr;
        Expr.BinOp2 <- x:Expr _ op:(Mul:`*` | Div:`/`) _ y:Expr;
        Expr.BinOp3 <- x:Expr op:#Alt[Pow:`^`] y:Expr;
        Expr.Paren <- `(` x:Expr[pr=*] `)`;
    }
    \end{verbatim}
    In addition to supplementary directives (e.g.,
    `\texttt{main}', `\texttt{prec}', `\texttt{prop}'),
    the parser stanza consists of a series of declarative, BNF-style {\em rules}
    of the form \texttt{X <- e},
    where \texttt{X} is the name of the rule (possibly consisting of multiple components,
    with dots), and \texttt{e} is the definition of the rule.
    Note that rules of the form \texttt{X.A}, \texttt{X.B}, and \texttt{X.C.D}
    all indicate the same nonterminal \texttt{X} for the left-hand side
    of the resulting context-free grammar rules,
    but in the generated ASTs, \texttt{X::A}, \texttt{X::B}, and \texttt{X::C::D}
    will be subclasses of the sum type \texttt{X}.

    We mention a number of additional features of the parser stanza:
    \begin{itemize}
        \item The \texttt{main} sub-stanza indicates a set of nonterminals
            that can be parsed at the top level,
            i.e., for which we can call \texttt{parse}
            in the generated API.
            Of these, the first one listed is the default
            if no nonterminal is specified at parsing time.
        \item The \texttt{prec} sub-stanza indicates a series of {\em precedence levels}
            for the rules.
            The details of precedence are detailed more fully in the original
            report~\cite[Sections 1.4, 3.5]{Zim22}.
            Note that unlike tools such as \texttt{yacc},
            our precedence spec operates at the level of rules,
            rather than at the level of operators.
            Concretely, it is implemented in terms of attribute constraints~\cite[Section 3.5]{Zim22}.
            We also note that the subexpression \texttt{Expr[pr=*]} in the \texttt{Expr.Paren}
            rule indicates an expression of arbitrary precedence,
            overriding what would otherwise be a highest-precedence constraint.
        \item The \texttt{prop} sub-stanza determines the configuration of \texttt{langcc}
            when processing the language definition.
            In particular, \texttt{prop \{ name\_strict; \}}
            indicates that every subexpression that corresponds to a field or a sum case
            in the AST must have a name (e.g., in `\texttt{x:Expr}', `\texttt{x}' is the name).
            Without the \texttt{name\_strict} annotation,
            fields may be unnamed,
            which will cause \texttt{langcc} to automatically generate names
            in the resulting AST---this is still valid,
            but makes it more difficult to write code against the generated API.
        \item The annotation \texttt{[I]} on \texttt{Expr.Id[I]} and \texttt{x:Expr[I]}
            indicates a boolean-valued {\em attribute} named \texttt{I}~\cite[Sections 1.4, 3.5]{Zim22}.
            Specifically, this means that only \texttt{Expr} instances arising from the \texttt{Expr.Id}
            rule will have the attribute \texttt{I},
            and conversely, in the \texttt{Stmt.Assign} rule,
            only \texttt{Expr} nonterminals which bear this attribute are valid
            at the indicated point in the right-hand side.
            Attributes can also be constrained via a standalone \texttt{attr} sub-stanza,
            similar to the \texttt{prec} sub-stanza
            (see \texttt{grammars/go.lang} in the source repository for examples).
            When attributes are specified inline, however (as in the example above),
            we adopt the general convention that an attribute mentioned
            in the right-hand side of a rule is a requirement,
            and an attribute mentioned in the left-hand side
            is a declaration that the attribute is satisfied.
            Further details can be found in~\cite[Section 3.5]{Zim22}.
        \item Many other features are available in parser BNF expressions, e.g.:
        \begin{itemize}
            \item A literal string may be {\em passed over} in the parser, written \texttt{@(`contents`)}.
                At parse time, this subexpression is ignored,
                but at pretty-print time, the contents of the string are inserted.
                This is often used for formatting strings such as \texttt{@(` `)} (a space)
                or \texttt{@(`\textbackslash n`)} (a newline);
                in fact, the former is so common that we have the special notation \texttt{\_}
                (as used in the example above) to denote \texttt{@(` `)}.
            \item The special expression \texttt{eps} may be used to denote the empty concatenation.
            \item The special expression \texttt{\#Alt(e)} may be used to denote
                the singleton alternation. To the parser, this is equivalent to the
                expression \texttt{e},
                but it may be important semantically for the generated AST.
            \item The expression \texttt{e*} may be used to denote a repeated expression \texttt{e}.
                In the generated AST, this automatically produces a vector.
            \item The expression \texttt{e+} may be used to denote a nonzero-repeated expression \texttt{e}.
                In the generated AST, this automatically produces a vector.
            \item The expression \texttt{\#L[e::delim]} may be used to denote a list of \texttt{e},
                delimited by \texttt{delim}.
                In the generated AST, this automatically produces a vector.
                In addition, there are several variants of the list expression:
                \begin{itemize}
                    \item \texttt{\#L[e::delim::]}, a list with a trailing delimiter.
                    \item \texttt{\#L[e::delim:?]}, a list with an optional trailing delimiter.
                    \item \texttt{\#L[e::+delim]}, a list with at least one element.
                    \item \texttt{\#L[e::++delim]}, a list with at least two elements.
                    \item \texttt{\#B[e::delim]}, a list which renders as an indented block in the pretty-printer.
                    \item \texttt{\#B2[e::delim]}, a list which renders as an indented double-spaced block
                        in the pretty-printer.
                    \item \texttt{\#T[e::delim]}, a list which renders as a top-level block in the pretty-printer.
                    \item \texttt{\#T2[e::delim]}, a list which renders as a top-level double-spaced block
                        in the pretty-printer.
                \end{itemize}
            \item The expression \texttt{e?} may be used to denote an optional expression \texttt{e};
                in the generated AST, this automatically produces an option type or a boolean,
                as appropriate.
            \item The expression \texttt{$\sim$X} may be used to indicate that the nonterminal \texttt{X}
                should be {\em unfolded}, i.e., that its beginning does not need to be predicted
                in recursive-descent style.
                If all nonterminals in the right-hand side are unfolded,
                this results in LR-style parsing (shift/reduce); if no nonterminals are unfolded,
                it results in RD-style parsing (shift/reduce/recur/ret).
                For further details, see~\cite[Section 3.6]{Zim22}.
        \end{itemize}
    \end{itemize}

\item \texttt{compile\_test} (optional): Describes a series of compilation tests to be performed when \texttt{langcc}
    compiles the language.
    Compilation tests are of the form \texttt{LR(k)} (resp., \texttt{!LR(k)}), where \texttt{k}
    is a nonnegative integer, indicating that LR compilation should succeed (resp., fail due to conflicts)
    for the given value of \texttt{k}.

\item \texttt{test} (optional): Describes a series of parsing tests to be performed when \texttt{langcc}
    compiles the language.
    Parsing tests are of the form \texttt{`abc`;} (resp., \texttt{`abc\#\#def;`}),
    indicating that parsing should succeed (resp., fail at the position indicated by \texttt{\#\#})
    on the given string.
    In addition, by default, parsing tests will verify that the pretty-printer outputs the same
    string as was parsed.
    If this is not desired for a given example, one may append the special marker \texttt{<<>>},
    as in \texttt{`contents` <<>>;}.

\end{itemize}

\section{Conflict tracing}
The underlying parsing theory~\cite{Zim22}
indicates that if \texttt{langcc} successfully compiles a language, then the behavior
of the parser is fully determined---though it may produce a parse error on strings that are
not in the language,
it will not fail by virtue of ambiguity.
However, not all grammars reach this point; some have {\em LR conflicts},
which \texttt{langcc} will report if it fails to compile a language for this reason.
When reporting LR conflicts, \texttt{langcc} endeavors to produce an exemplar,
a short ``confusing input pair'', that explains the conflict
in a way that is much more intuitive than the opaque shift/reduce errors produced by
tools such as \texttt{yacc}.
Further details of the conflict tracing procedure are described in \cite[Section 3.4]{Zim22};
here we only provide a basic example.
The following is a conflict that arises if we attempt to run \texttt{langcc}
on a simple arithmetic expression language, without using the appropriate precedence declarations.
\newpage
\begin{verbatim}
    ===== LR conflict 1 of 2

                &Expr                           &Expr    
                                      RecurStep(Expr)    
                 Expr                              id    
       X0=(`+` | `-`)                             `+`    
                 Expr                              id    
                                                         
                         Reduce(Expr -> Expr X0 Expr)    Shift
                                                         
                                                  `+`    `+`
                                                   id    id    
\end{verbatim}
On the left side, we see the expression that produced the conflicting prefix
(\texttt{Expr X0 Expr}),
and on the right side, a concrete input that might correspond to this expression
(\texttt{id `+` id}).
Below this are the two conflicting actions, (1) \texttt{Reduce} by the indicated production
\texttt{Expr -> Expr X0 Expr}
and (2) \texttt{Shift} the next token.
Finally, there is a completion of the ``confusing input pair'',
showing that either action is possible with the given $k=1$ lookahead (\texttt{`+`}).
In this example, the language is actually ambiguous;
one cannot decide whether to parse \texttt{id+id+id} as \texttt{(id+id)+id} or as \texttt{id+(id+id)}.
However, a language need not be ambiguous in order to present LR conflicts---in general,
all that is required is that the parser is unable to determine what LR action to take
with $k$~tokens of lookahead.

\section{Output API}

When a language \texttt{X.lang} is successfully compiled, \texttt{langcc} outputs
files \texttt{gen\_path/X\_\_gen.hpp} and \texttt{gen\_path/X\_\_gen.cpp},
which contain the AST definitions, lexer, parser,
and pretty-printer, as well as miscellaneous utilities such as debug-printers and hashing functions.
The following is a basic example of how to use the generated API (from \texttt{examples/calc}):
\begin{verbatim}
    auto L = lang::calc::init();
    auto Q = L->quote_env();
    unordered_map<string, Int> env;
    string l;

    while (true) {
        getline(cin, l);
        if (!cin.good()) {
            return 0;
        }

        auto gen = make_rc<Gensym>();
        auto parse = L->parse_ext(
            vec_from_std_string(l), None<string>(), gen, nullptr);

        if (!parse->is_success()) {
            LG_ERR("\nParse error: {}\n", parse->err_.as_some());
            continue;
        }

        auto stmt = parse->res_.as_some()->as_Stmt();
        try {
            fmt(cerr, "{}\n", stmt_eval(stmt, env));
        } catch (const CalcError& err) {
            LG_ERR("\nError: {}\n{}",
                err.desc_,
                parse->lex_->location_fmt_str(err.e_blame_->bounds_));
            continue;
        }
    }
\end{verbatim}
Note that the generated parsing procedure returns a structure \texttt{parse},
and if \texttt{!parse->is\_success()},
then \texttt{parse->err\_} contains the parse error;
when formatted,
it can be printed in user-readable form, resulting in a message like the following:
\begin{verbatim}
    Parse error: Unexpected token: `/`
    Line 1, column 10:
    
      7 + (5 + / 3)
               ^    
\end{verbatim}
If, on the other hand, parsing succeeds, then the resulting AST element
is given by \texttt{parse->res\_}.
By default, it is a generic \texttt{lang::calc::Node\_T} (a sum type),
but in this case, the toplevel parse is known to be a \texttt{Stmt},
and we downcast it with \texttt{as\_Stmt()},
then call our function \texttt{stmt\_eval}, which in turn calls \texttt{expr\_eval}
on constituent expressions.
The following is excerpted from \texttt{expr\_eval}:
\begin{verbatim}
    Int expr_eval(Expr_T e, const unordered_map<string, Int>& env) {
        if (e->is_Lit()) {
            auto val_str = e->as_Lit()->as_Int_()->val_.to_std_string();
            return string_to_int(val_str).as_some();
        // ...
        } else if (e->is_BinOp2()) {
            auto cc = e->as_BinOp2();
            auto xr = expr_eval(cc->x_, env);
            auto yr = expr_eval(cc->y_, env);
            if (cc->op_->is_Mul()) {
                return xr * yr;
            } else if (cc->op_->is_Div()) {
                if (yr == 0) {
                    throw CalcError(fmt_str("Division by zero"), cc->op_);
                }
                return xr / yr;
            } else {
                AX();
            }
        // ...
\end{verbatim}
As this example shows, it is straightforward to decompose the AST sum types
to obtain the values enclosed.
We mention one additional feature:~note that in the case of division by zero,
we throw an error that includes the syntax element \texttt{cc->op\_}.
In fact, this syntax element carries with it its position in the input,
which enables us to produce, at top level, error messages of the following form:
\begin{verbatim}
    Error: Division by zero
    Line 1, column 3:

      4 / (3 - (15 / 5))
        ^                
\end{verbatim}
indicating precisely which division triggered the error.
Evidently, this functionality easily generalizes to source-position error reporting
in more complex languages.

\section{Bootstrapping}

We finally mention one additional property of \texttt{langcc}:~the class of grammars
is general enough that the tool is {\em self-hosting}---that is, one can express the ``language
of languages'' in the ``language of languages'' itself, and use \texttt{langcc}
to generate its own compiler front-end. In fact, we do this in the canonical
implementation; see the files \texttt{bootstrap.sh} and \texttt{grammars/meta.lang}
in the source repository for more details.
The metalanguage is surprisingly concise, requiring only 189 lines of code.
The following is a brief excerpt:
\begin{verbatim}
    ParseExpr.Pass <- `@` `(` s:str_lit `)`;
    ParseExpr.Paren <- `(` x:ParseExpr[pr=*] `)`;
    ParseExpr.Name <- name:IdBase `:` e:ParseExpr;
    ParseExpr.List <- ty:~ParseExprListType
        `[`
        elem:ParseExpr[pr=*]
        num:ParseExprListNum
        delim:ParseExpr[pr=*]
        end_delim:(NONE:eps | OPTIONAL:`:?` | SOME:`::`)
        `]`;
    ParseExpr.Unfold <- `~` e:ParseExpr;
    ParseExpr.AttrReq <- e:ParseExpr `[` attrs:#L[AttrReq::`,`_] `]`;

        AttrReq.Base <- k:IdBase;
        AttrReq.PrecStar <- `pr` `=` `*`;

        ParseExprListType.List <- `#L`;
        ParseExprListType.Block <- `#B`;
        // ...
\end{verbatim}
We note that the syntax is very compact,
and corresponds to little more than one would write on the whiteboard for an informal BNF grammar.

\bibliographystyle{alpha}
\bibliography{langcc}

\begin{thebibliography}{Zim22}

\bibitem[Zim22]{Zim22}
Joe Zimmerman.
\newblock Practical {LR} parser generation.
\newblock {\em arXiv}, 2022.

\end{thebibliography}

\end{document}